\documentclass[aps,prc,showpacs,nofootinbib]{revtex4}

\usepackage{epsfig}
\usepackage{graphicx}

\begin{document}

\title{Polarization effects in the reaction $\bar p+p\rightarrow e^++e^-$ in presence of two--photon exchange}

\author{G. I. Gakh \footnote{e-mail: \it gakh@kipt.kharkov.ua}}  
 
\affiliation{ National Science Centre "Kharkov Institute of Physics and Technology", 61108 Akademicheskaya 1, Kharkov,
Ukraine} 
\author{E. Tomasi--Gustafsson}
\affiliation{\it DAPNIA/SPhN, CEA/Saclay, 91191 Gif-sur-Yvette Cedex,
France }



\begin{abstract}
Polarization observables for the reaction $\bar p+p\rightarrow e^++e^-$ are given in terms of three independent complex amplitudes, in presence of two photon exchange. General expressions for the 
differential cross section and the  polarization observables are given and  model independent properties are derived. Polarization 
effects depending on the polarization of the antiproton beam, the target and of the electron in the final state, have been calculated.
\end{abstract}

\maketitle
\section{Introduction}

The problem of the presence of the two--photon--exchange (TPE) contribution
for  elastic electron--proton scattering at relatively large momentum
transfer is very actual. Intensive theoretical and experimental activity is
under way, related, in particular, to the discrepancy between the experimental
results on the proton electromagnetic form factors (FFs), extracted by
different procedures, through the Rosenbluth fit \cite{An94} or from the
polarization transfer method \cite{JG00} (for a recent discussion see Ref.
\cite{Ar04a}).

Estimations of the TPE contribution to the elastic electron--deuteron
scattering were firstly discussed in Refs. \cite{Gu73, Fr73} in the
framework of the Glauber theory. It was shown \cite{Gu73} that this
contribution decreases very slowly with momentum transfer squared $q^2$ and
may dominate the cross section at high $q^2$ values. Since the TPE amplitude
is essentially imaginary, the difference between positron and electron
scattering cross sections depends upon the small real part of the TPE
amplitude \cite{Gu73}. Recoil polarization effects may be substantial, in
the region where the one-- and two--photon--exchange contributions are
comparable. If the TPE mechanism becomes sizeable, the straightforward
extraction  of FFs from the experimental data is no longer possible
\cite{Gu73}.

It is known that double scattering dominates in collisions of high--energy
hadrons with deuterons at high $q^2$ values, and in Ref. \cite{Fr73} it was
predicted that the TPE effect in  elastic electron--deuteron scattering
should represent 10\%  effect at $q^2\cong$ 1.3 GeV$^2.$ At the same time
the importance of the TPE mechanism was considered in Ref. \cite{Bo73}.
The fact that the TPE mechanism, where the momentum transfer is shared
between the two virtual photons, can become important with increasing $q^2$
value was already indicated more than thirty years ago \cite{Gu73,Fr73,Bo73}.

This mechanism was never directly observed in an experiment, but recent measurements of the asymmetry in the scattering of transversely polarized electrons on unpolarized protons, give values different from zero, contrary to what is expected in the Born approximation \cite{We00,Ma04}. This observable is related to the imaginary part of the interference between one and two photon exchange and can be related only indirectly to the real part of the interference, which plays a role in the elastic $ep$ cross section.

Measurements of the ratio of the electric to the magnetic proton FFs,  $G_E/G_M$, have been performed at JLab in polarized $ep$ elastic  scattering, ${\vec e}+p\to e+{\vec p}$  \cite{JG00}. The transverse, $P_t$, and the  longitudinal, $P_l$, components of the recoil proton 
polarization in the electron scattering plane are directly related to the ratio of the electromagnetic proton  FFs. This method, firstly suggested in Ref. \cite{Re68}, could be applied only recently, due to the availability of high intensity, high polarized electron beams, hadron polarimeters in the GeV range and large acceptance spectrometers. 

The data \cite{JG00} have been obtained in the region 0.3 GeV$^2$
$\leq Q^2 \leq$ 5.6 GeV$^2$,  and reveal a remarkable fall of the ratio
$G_E/G_M$ when $Q^2$ increases, in disagreement with the data obtained
by the Rosenbluth technique, which show that this ratio is constant.

In Ref. \cite{Du04} it has been shown, on the basis of a VMD inspired model
taking into account ten resonances, that the polarization data  \cite{JG00}
may be consistent with all known FF properties, including also QCD asymptotics
and that $G_E$ will  vanish around $q^2$=-15 GeV$^2$. A zero, and
eventually negative values of $G_E$, if confirmed by the planned
experiment \cite{04108}, will seriously constrain the nucleon models.

From the theoretical point of view, it seems unavoidable to consider the
problem of the TPE contribution in the $\bar p+p\to e^++e^-$ reaction. The
process $\bar p+p\to e^++e^-$ and its crossing channel, $e+p\to e+p$, must
have common mechanisms. The process $\bar p + p\to e^++ e^-$ is very
convenient to study the polarization effects induced by collisions of
polarized protons and antiprotons, but the measurement of the final--lepton
polarization cannot be considered as a realizable experiment. However, for
completeness, we will also calculate observables related to the final
electron polarization.

The TPE contribution in the $\bar p+p\to e^++e^-$ reaction results, first of
all, in a nonlocal spin structure of the matrix element. This makes the
analysis of polarization effects more complicated with respect to the case of
the one--photon--exchange mechanism. Such analysis can be done similarly to
the case of hadronic reactions among spin 1/2 particles, such as, for example,
$n+p\to n+ p$ scattering \cite{Go57}.

At our knowledge, the annihilation reaction $\bar p + p\to \ell^++\ell^-$,
$\ell=e$ or $\mu $  was firstly considered in Ref. \cite{Zi62} in the case
of unpolarized particles, where the differential cross section was calculated
both in the center of mass (CMS) and in the laboratory (Lab) systems. As
already mentioned, if nucleon FFs decrease rapidly in  time--like region,
then, just as in space--like region, it is possible that the TPE mechanism
becomes important.

The general case of polarized initial particles (antiproton beam or/and
proton target) in $\bar p + p\to e^++e^-$ has been firstly investigated in
Ref. \cite{Bi93}, with particular attention to the determination of the
phases of FFs, and more recently in Ref. \cite{ETG05}. The relations between
the measurable asymmetries in terms of the electromagnetic FFs, $G_M$ and
$G_E$, in the time--like region were derived, assuming one photon exchange.

In this paper we consider the reaction
\begin{equation}\label{eq:eq1}
\bar p+p\to e^++e^-.
\end{equation}
We derive here the expressions for the differential cross section and
various polarization observables for the case when the matrix element
contains the TPE contribution. The parametrization of the TPE term is done
following the analytic continuation to the time--like region of the approach
used (in the space--like region) in Refs. \cite{Gu03,Re04a,Re04b,Re04c,Ch04}.
Using some approximations or in framework of a model, it was shown TPE could  account, at least partially, for the apparent discrepancy between the Rosenbluth and the polarization transfer methods. 

Another approach is taken in Refs. \cite{Re04a,Re04b,Re04c}, where the
purpose is to derive general expressions for the polarization observables in
the elastic electron--nucleon scattering and to suggest model independent
methods to extract nucleon electromagnetic FFs even in presence of the TPE
contribution (parametrized in the tensor \cite{Re04a} or axial \cite{Re04b}
forms), without underlying assumptions. We use the tensor form of the TPE
contribution parametrization, and follow the approach of Ref. \cite{Re04b}.

\section{Differential cross section}

Let us consider the process (\ref{eq:eq1}) in the general case of polarized  beam
and target and measuring the polarization of the outgoing electron. The
starting point of our analysis of the reaction (\ref{eq:eq1}) is the following general
parametrization of the spin structure of the matrix element for this reaction,
according to the approach used in Refs. \cite{Gu03,Re04a,Re04b,Re04c,Ch04}
\begin{equation}\label{eq:eq2}
{\it M} = -\frac{e^2}{q^2}j_{\mu }J_{\mu }, \mbox{~with~} j_{\mu }=\bar u(k_2)\gamma _{\mu }u(-k_1),
\end{equation}
and
$$
J_{\mu }=\bar u(-p_2)[\tilde G_M(q^2, t)\gamma_{\mu }+\frac{P_{\mu }}{m}
\tilde F_2(q^2, t)+\frac{P_{\mu }}{m^2}\hat K F_3(q^2, t)]u(p_1),$$
where $K=(k_1-k_2)/2,~P=(p_2-p_1)/2,$ $p_1~(p_2)$ and $k_1~(k_2)$ are the 
four--momenta of proton (antiproton) and positron (electron), respectively; 
$q^2=(p_1+p_2)^2,$~ $t=K\cdot P$, $m $ is the nucleon mass, $\tilde G_M,$ 
$\tilde F_2$ and $ F_3$ are  complex functions of two independent 
variables $q^2$ and $t.$

The spin structure of the matrix element of the electron--nucleon scattering (\ref{eq:eq2})
can be derived in analogy with the elastic neutron--proton 
scattering \cite{Go57} assuming  general properties of the electron--hadron
interaction, such as P--invariance and  relativistic invariance. 
Taking into account the identity of the initial and final states and the 
T--invariance of the electromagnetic interaction, the process of the electron--
nucleon scattering is characterized by three invariant complex amplitudes 
(in the limit of zero electron mass). The spin structure of the matrix element 
for the $\bar p+p\rightarrow e^++e^-$ reaction is obtained from the matrix 
element of the elastic electron--nucleon scattering by analytic continuation. 

In the Born (one--photon--exchange) approximation we have
\begin{equation}\label{eq:eq3}
 \tilde G_M^{Born}(q^2, t)=G_M(q^2), \ 
 \tilde F_2^{Born}(q^2, t)=F_2(q^2), \
  F_3^{Born}(q^2, t)=0, \ 
\end{equation}
where $G_M(q^2)$ and $F_2(q^2)$ are the magnetic and Pauli proton
electromagnetic FFs, respectively, which are complex functions 
of the variable $q^2.$ The complex nature of FFs in time-like region is due to the strong interaction 
between proton and antiproton in the initial state. 

In the following we use 
the Sachs magnetic $G_M(q^2)$ and charge $G_E(q^2)$ proton FFs 
which are related to the Dirac proton FF $F_1(q^2)$ and to $F_2(q^2)$ as follows
\begin{equation}\label{eq:eq4}
 G_M=F_1+F_2, \  G_E=F_1+\tau F_2, \  \tau =\frac{q^2}{4m^2}.
\end{equation}

To disentangle the effects of the Born and TPE contributions, let us single 
out the dominant contribution and define the following decompositions of the 
amplitudes:  
\begin{equation}\label{eq:eq5}
\tilde G_M(q^2, t)=G_M(q^2)+\Delta G_M(q^2, t), \
\tilde F_2(q^2, t)=F_2(q^2)+\Delta F_2(q^2, t). \
\end{equation}
Instead of the amplitude $\tilde F_2$ we use the linear combination
\begin{equation}\label{eq:eq6}
\tilde G_E(q^2, t)=G_E(q^2)+\Delta G_E(q^2, t). \
\end{equation}
We neglect below the bilinear combinations of the terms $\Delta G_M,$~$\Delta G_E$ and $F_3$ since they are smaller (at least of the order of $\alpha $), in 
comparison with the dominant ones.

Then the differential cross section of the reaction (\ref{eq:eq1}) can be written in CMS as follows: 
\begin{equation}\label{eq:eq7}
\frac{d\sigma}{d\Omega}=\frac{\alpha ^2}{4q^6}\frac{E}{p}L_{\mu\nu }
H_{\mu\nu },~
L_{\mu\nu }=j_{\mu }j^*_{\nu},~ 
H_{\mu\nu }=J_{\mu }J^*_{\nu }, 
\end{equation}
where $E (p)$ is the energy (momentum) of the antiproton. In the case of longitudinally polarized electrons the leptonic tensor has the form  
\begin{equation}\label{eq:eq8}
L_{\mu\nu }=-q^2g_{\mu\nu }+2(k_{1\mu }k_{2\nu }+k_{1\nu }k_{2\mu })+
2i<\mu\nu qk_2>, \
\end{equation}
where $<\mu\nu ab>=\varepsilon_{\mu\nu\rho\sigma }a_{\rho}b_{\sigma}$. Other components of the electron polarization lead to a suppression by a factor $m_e/m$.

Taking into account the polarization states of the beam and target, the 
hadronic tensor can be written as the sum of three tensors as follows:
\begin{equation}\label{eq:eq9}
H_{\mu\nu }=H_{\mu\nu }^{(0)}+H_{\mu\nu }^{(1)}+H_{\mu\nu }^{(2)}, \
\end{equation}
where the tensor $H_{\mu\nu }^{(0)}$ corresponds to the unpolarized beam and target, the tensor $H_{\mu\nu }^{(1)}$ describes the production of $e^+e^-$ by 
polarized beam or target and the tensor $H_{\mu\nu }^{(2)}$ corresponds to polarized beam and polarized target.

Since the presence of the TPE contribution leads to the term of the hadronic
current which contains the momenta from the leptonic vertex, the general structure of
the $H_{\mu\nu }^{(0)}$ tensor becomes more complicated: instead of the two standard  
structure functions we have five ones (as in the case of 
$\gamma ^*d\rightarrow np$ or $\gamma ^* N\rightarrow \pi N$). So, 
the general structure of this tensor can be written as 
\begin{equation}\label{eq:eq10}
H_{\mu\nu }^{(0)}=H_1\tilde g_{\mu\nu }+H_2P_{\mu }P_{\nu }
+H_3K_{\mu }K_{\nu }+H_4(K_{\mu }P_{\nu }+K_{\nu }P_{\mu})+
iH_5(K_{\mu }P_{\nu }-K_{\nu }P_{\mu}), 
\end{equation}
where $\tilde g_{\mu\nu }=g_{\mu\nu }-q_{\mu }q_{\nu}/q^2.$ One gets the
following expressions for these structure functions for the case of the hadronic
current given by Eq. (\ref{eq:eq2}):
\begin{eqnarray}
H_1&=&-2q^2(|G_M|^2+2ReG_M\Delta G_M^*), \nonumber\\ 
H_2&=&\frac{8}{\tau -1}\biggl [|G_E|^2-\tau |G_M|^2+2ReG_E\Delta G_E^*-
2\tau ReG_M\Delta G_M^*+
\nonumber\\ 
&&2\sqrt{\tau (\tau -1)}\cos\theta 
Re(G_E-\tau G_M)F_3^*\biggr ], \nonumber\\ 
H_3&=&0, ~ H_4=-8\tau ReG_MF_3^*, ~H_5=-8\tau ImG_MF_3^*,
\label{eq:eq11} 
\end{eqnarray}
where $\theta $ is the angle between the electron and the antiproton momenta in the 
$\bar p+p\rightarrow e^++e^-$ reaction CMS. One can see that the structure
functions $H_4$ and $H_5$ are completely determined by the TPE terms: in the 
absence of these terms we have the standard tensor structure for $H_{\mu\nu }^{(0)}$.

The differential cross section of the reaction (\ref{eq:eq1}) for the case of unpolarized 
particles has the form: 
\begin{equation}\label{eq:eq12}
\frac{d\sigma}{d\Omega } = \frac{\alpha^2}{4q^2}\sqrt{\frac{\tau }
{\tau -1}}D,
\end{equation}
\begin{eqnarray*}
D&=&(1+\cos^2\theta )(|G_M|^2+2ReG_M\Delta G_M^*)+\frac{1}{\tau }
\sin^2\theta (|G_E|^2+2ReG_E\Delta G_E^*)+ \nonumber\\ 
&&2\sqrt{\tau (\tau -1)}\cos\theta \sin^2\theta Re(\frac{1}{\tau }G_E-G_M)F_3^*.
\end{eqnarray*}
One can see that in the Born approximation the expression (\ref{eq:eq12}) reduces to the
result obtained in Refs. \cite{Bi93,ETG05}. The contribution of the one--photon--exchange diagram leads to an even function of $\cos\theta $, whereas the TPE contribution leads to four new terms of the order of $\alpha $  compared to the dominant contribution. 

At the reaction threshold where $q^2 = 4m^2$, one gets $G_M=G_E$ 
and the differential cross section becomes $\theta-$independent in the Born
approximation. This is not anymore true in presence of TPE terms.

As it was pointed out in Ref. \cite{Pu61}, for the processes of the type $e^++e^-\to h^++h^-$, except in particular cases, 
the term of the cross section due to TPE is an odd function of the variable $\cos\theta $. Therefore,
it does not contribute to the differential cross section for $\theta= 90^0.$

\section{Single spin Polarization observables}

Let us consider the case when the antiproton beam is polarized. Then, if the
hadronic current is given by Eq. (\ref{eq:eq2}), the hadronic tensor $H_{\mu\nu }^{(1)}$ can be written as:
\begin{eqnarray}
H_{\mu\nu }^{(1)}&=&-\frac{2i}{m}\biggl [m^2|\tilde G_M|^2<\mu\nu qs_2>+
(\tau -1)^{-1}Re\tilde G_M(\tilde G_E-\tilde G_M)^*(<\mu p_1p_2s_2>P_{\nu }-\nonumber\\
&&<\nu p_1p_2s_2>P_{\mu })+ReG_MF_3^*(<\mu kqs_2>P_{\nu }-
<\nu kqs_2>P_{\mu})\biggr ]+ \nonumber\\
&&\frac{2}{m(\tau -1)}\biggl [Im\tilde G_M\tilde G_E^*(<\mu p_1p_2s_2>P_{\nu }+
<\nu p_1p_2s_2>P_{\mu })+ \nonumber\\
&&(\tau -1)ImG_MF_3^*(<\mu kqs_2>P_{\nu }+<\nu kqs_2>P_{\mu})- \nonumber\\
&&
\frac{2}{m^2}<s_2p_2p_1k>Im(G_E-G_M)F_3^*P_{\mu }P_{\nu }\biggr ],\label{eq:eq13} 
\end{eqnarray}
where $s_{2\mu }$ is the antiproton polarization four--vector ($p_2\cdot s_2 = 0$). 

Note that, unlike the elastic electron--nucleon scattering in the Born approximation, the hadronic tensor in the time--like region contains a  symmetric part even in the Born
approximation due to the fact that nucleon FFs are complex. Taking into
account the TPE contribution leads to additional terms in the symmetric
part of this tensor.

The polarization four--vector of a relativistic particle, $s_{\mu }$, in a reference  system where its momentum, 
${\vec p}$, is connected with the polarization vector, ${\vec \chi}$, in its rest frame by a Lorentz boost is:
$${\vec s}={\vec \chi}+\frac{{\vec p}\cdot {\vec \chi}{\vec p}}{m(E+p)}, \ 
s^0=\frac{1}{m}{\vec p}\cdot {\vec s}. $$ 
Let us define a coordinate frame in CMS of the reaction (\ref{eq:eq1}), where the $z$ axis is
directed along the antiproton momentum ${\vec p}$, the $y$ axis is directed along
the vector ${\vec p}\times {\vec k}$, (${\vec k}$ is the electron momentum),
and the $x$ axis forms a left--handed coordinate system. In this frame the
components of the unit vectors are: $\hat {\vec p}=(0,0,1)$ and 
$\hat {\vec k}=(\sin\theta ,0,\cos\theta)$ with 
$\hat {\vec p}\cdot \hat {\vec k}=\cos\theta.$

The presence of a symmetrical part in the hadronic tensor (\ref{eq:eq13}) leads to a 
non--zero single--spin asymmetry which can be written as
\begin{equation}\label{eq:eq14}
A_y(\theta )=\frac{2\sin\theta }{\sqrt{\tau }D}\biggl [\cos\theta ~Im(
G_MG_E^*+G_M\Delta G_E^*-G_E\Delta G_M^*)+
\end{equation}
$$+\sqrt{\tau (\tau -1)}Im(\cos^2\theta~ G_M+\sin^2\theta~ G_E)F_3^*\biggr ]. $$

Again, in the Born approximation this expression reduces to the result of Ref.
\cite{Bi93}. One can see that:

- $A_y(\theta )$ is determined by the spin vector component which is
perpendicular to the reaction plane;

- $A_y(\theta )$,  being a T--odd quantity, does not vanish even in the
one--photon--exchange approximation due to the complex nature of the nucleon 
FFs in the time--like region. This is the principal difference with the elastic
electron--nucleon scattering.

Let us consider two particular kinematical cases: 

- when the electron is scattered at $\theta = 90^0 $. 

- the reaction threshold.

For  $\theta = 90^0$, in the Born approximation $A_y(\theta )$ vanishes. The presence of the TPE contributions leads to
a non--zero value of $A_y(\theta )$  at $\theta = 90^0 $ and this value is given by a
simple expression
$$A_y(90^0)=2\frac{\sqrt{\tau -1}}{\overline  D}ImG_EF_3^*, \ \ 
\overline  D=D(\theta =90^0). $$
This quantity is expected to be small due to the fact that it is determined by the 
interference of the one--photon and two--photon exchange amplitudes and is 
of the order of $\alpha $. One can see that this asymmetry is an increasing function of the
variable $q^2$: this is due to the presence of the kinematical factor containing ${\tau }$  and to the steep decreasing of the nucleon FFs with $q^2$ while the TPE mechanism becomes more important when  $q^2$ increases. So, the measurement of this asymmetry at $\theta = 90^0$ can
give information about the TPE contribution and its behaviour as a function of
$q^2.$

At threshold, in the Born approximation,  $A_y^{th}(\theta )$ has to vanish, due to the relation $G_E=G_M$. Including the TPE contributions, the asymmetry becomes: 
$$A_y^{th}(\theta )=\frac{\sin2\theta }{D_{th}}ImG_M(\Delta G_E-\Delta G_M)^*. $$
Note that, at threshold, this asymmetry can be equal to zero, if $\Delta G_E=\Delta G_M$. In this case the differential cross section does not contain any explicit dependence on the angular variable $\theta $, but only through the amplitudes $\Delta G_{E,M}$ which, in
the general case, depend on the variable $\theta $.

The importance of the TPE contributions in $A_y^{th}(\theta )$ at an arbitrary
scattering angle will increase as $q^2$ increases. This is due to the presence
of the kinematical factor containing $\tau$ and it is expected that the TPE amplitudes decrease more slowly with $q^2$ compared with the nucleon FFs.

The antisymmetrical part of the hadronic tensor $H_{\mu\nu }^{(0)}$ leads to another single--spin observable: the final electron gets  a
transverse polarization (orthogonal to the reaction plane) in the annihilation
of unpolarized proton and antiproton. The expression for this polarization is: 
$$P_y^{(e)}(\theta )=2\frac{m_e}{m}\frac{\sqrt{\tau -1}}{D}\sin\theta ImG_MF_3^*, \\ $$
where $m_e$ is the electron mass. One can see that

- $P_y^{(e)}$ has a T--odd nature, since it is determined by the imaginary
part of the product of $G_M$ and of the amplitude $F_3.$

- $P_y^{(e)}$  is entirely due to the TPE mechanism and it vanishes in the Born 
approximation.

- Since it is a transverse polarization, it is suppressed by a factor $(m_e/m).$
The polarization for the case of production of $\mu ^+\mu ^-$-pair is
essentially  larger ($m_{\mu }/m_e$=200) and for $\tau ^+\tau ^-$ production 
one finds no additional suppression. Another advantage of detecting heavy leptons is that the polarization of unstable particles ($\mu $ and $\tau $) can be measured through the angular distribution of their decay products.  

- $P_y^{(e)}$  vanishes at threshold, also in presence of TPE contribution.

- $P_y^{(e)}$  increases when $q^2$ becomes larger. The reasons are the
same as for the asymmetry $A_y$ (see the discussion above).

Let us consider now the polarization transfer when the antiproton beam
is polarized and the polarization of the produced electron is measured. We consider only the longitudinal polarization of
the final electron because in this case the suppression factor $m_e/m$ is
absent. The corresponding observables are: 
\begin{eqnarray}
A_x&=&\frac{2\sin\theta }{\sqrt{\tau }D}\biggl [ReG_MG_E^*+Re(G_M\Delta G_E^*+
G_E\Delta G_M^*)+\sqrt{\tau (\tau -1)}\cos\theta ReG_MF_3^*\biggr ], 
\nonumber\\
A_z&=&\frac{2}{D}\biggl [\cos\theta (|G_M|^2+
2ReG_M\Delta G_M^*)-\sqrt{\tau (\tau -1)}\sin^2\theta ReG_MF_3^*\biggr ]. 
\label{eq:eq15}
\end{eqnarray}

These coefficients are T--even observables and they are nonzero in the 
Born approximation, and also for elastic electron--nucleon scattering. The 
coefficient $A_z$ vanishes at $\theta = 90^0$ in 
the Born approximation. But the presence of the TPE term $F_3$ in the
electromagnetic hadron current leads to a nonzero value of this quantity, driven by the term $ReG_MF_3^*.$

The expressions (\ref{eq:eq15}), in the one--photon--exchange approximation, coincide with
the results for the polarization vector components of the nucleon in the 
$e^++e^-\rightarrow N+\bar N$ reaction, when the electron beam is longitudinally
polarized \cite{Du96,Br04}.

\section{Double spin polarization observables}

Let us consider the case when the polarized antiproton beam annihilates 
with a polarized proton target. The corresponding hadronic tensor 
$H_{\mu\nu }^{(2)}$ can be written as: 
\begin{eqnarray}
H_{\mu\nu }^{(2)}&=&C_1g_{\mu\nu }+C_2P_{\mu }P_{\nu }+
C_3(P_{\mu }s_{1\nu }+P_{\nu }s_{1\mu})+C_4(P_{\mu }s_{2\nu }+
P_{\nu }s_{2\mu})+ \nonumber \\
&&C_5(s_{1\mu }s_{2\nu }+s_{1\nu }s_{2\mu})+
C_6(P_{\mu }K_{\nu }+P_{\nu }K_{\mu})+
iC_7(P_{\mu }s_{1\nu }-P_{\nu }s_{1\mu})+ \nonumber \\ 
&&
iC_8(P_{\mu }s_{2\nu }-P_{\nu }s_{2\mu²})+
iC_9(P_{\mu }K_{\nu }-P_{\nu }K_{\mu}),
\label{eq:eq16}
\end{eqnarray}
where $s_{1\mu }$ is the proton polarization four-vector ($p_1\cdot s_1 = 0$) and the terms proportional to $q_{\mu }$ or $q_{\nu }$ were omitted, since they do 
not contribute to the cross section and to the polarization observables (due to the 
conservation of the leptonic current). The structure functions have the 
following form
\begin{eqnarray}
C_1&=&\frac{1}{2}(q^2s_1\cdot s_2-2q\cdot s_1q\cdot s_2)|\tilde G_M|^2,\nonumber \\
C_2&=&\frac{2}{\tau -1}\biggl [\tau |\tilde G_M|^2-|\tilde G_E|^2+
2\frac{K\cdot P}{m^2}Re(\tau G_M-G_E)F_3^*\biggr ]s_1\cdot s_2+ \nonumber \\ 
&& \frac{q\cdot s_1q\cdot s_2}{m^2(\tau -1)^2}|\tilde G_E-\tilde G_M|^2+ 
\nonumber \\ 
&& 
\frac{2}{m^2(\tau -1)}(q\cdot s_1K\cdot s_2-q\cdot s_2K\cdot s_1)
Re(G_E-\tau G_M)F_3^*,
\nonumber \\ 
C_3&=&ReE_1,~C_4=ReE_2, ~C_6=ReE_3,~ 
C_5=-\frac{q^2}{2}|\tilde G_M|^2, \nonumber \\ 
E_1&=&\frac{q\cdot s_2}{\tau -1}(\tau |\tilde G_M|^2-\tilde G_E\tilde G_M^*)+
\frac{1}{2m^2}(2K\cdot Pq\cdot s_2-q^2K\cdot s_2)F_3G_M^*, \nonumber \\ 
E_2&=&-\frac{q\cdot s_1}{\tau -1}(\tau |\tilde G_M|^2-\tilde G_E\tilde G_M^*)-
\frac{1}{2m^2}(2K\cdot Pq\cdot s_1+q^2K\cdot s_1)F_3G_M^*, \nonumber \\ 
E_3&=&\frac{1}{2m^2}(q^2s_1\cdot s_2-2q\cdot s_1q\cdot s_2)F_3G_M^*,\nonumber \\
C_7&=&ImE_1,~C_8=ImE_2, ~C_9=ImE_3. 
\label{eq:eq16a}
\end{eqnarray}

The non--zero spin correlation coefficients between the polarizations of beam 
and target (when the final leptons are unpolarized) can be written as: 
\begin{eqnarray}
D_{xx}&=&\frac{\sin^2\theta }{D}\biggl [|G_M|^2+2ReG_M\Delta G_M^*+
\frac{1}{\tau }(|G_E|^2+2ReG_E\Delta G_E^*)+ \nonumber \\ 
&&2\sqrt{\tau (\tau -1)}\cos\theta Re(G_M+\frac{1}{\tau }G_E)F_3^*\biggr ], 
\nonumber \\ 
D_{yy}&=&\frac{\sin^2\theta }{D}\biggl [\frac{1}{\tau }(|G_E|^2+2ReG_E\Delta G_E^*)-
|G_M|^2-2ReG_M\Delta G_M^*- \nonumber \\ 
&&2\sqrt{\tau (\tau -1)}\cos\theta Re(G_M-\frac{1}{\tau }G_E)F_3^*\biggr ], \nonumber \\ D_{zz}&=&\frac{1}{D}\biggl [(1+\cos^2\theta )(|G_M|^2+2ReG_M\Delta G_M^*)-
\frac{1}{\tau }\sin^2\theta (|G_E|^2+2ReG_E\Delta G_E^*) -
\nonumber \\ 
&&2\sqrt{\tau (\tau -1)}\cos\theta \sin^2\theta 
Re(G_M+\frac{1}{\tau }G_E)F_3^*\biggr ], \nonumber \\ 
D_{xz}&=&D_{zx}=\frac{\sin2\theta }{\sqrt{\tau }D}\biggl [Re(G_MG_E^*+
G_M\Delta G_E^*+G_E\Delta G_M^*)+ 
\nonumber \\ 
&&\sqrt{\tau (\tau -1)}\cos\theta Re(
G_M-tan^2\theta G_E)F_3^*\biggr ]. 
\label{eq:eq17}
\end{eqnarray}
For completeness, we give here the nonzero coefficients for the case of a 
longitudinally polarized electron:
\begin{eqnarray}
D_{xy}=D_{yx}&=&\frac{1}{D}\sqrt{\tau (\tau -1)}\sin^2\theta ImG_MF_3^*,
\nonumber \\ 
D_{zy}=D_{yz}&=&\frac{\sin\theta }{\sqrt{\tau }D}Im \left (G_{Mp}
G_{Ep}^*+G_{Mp}\Delta G_{Ep}^*-G_{Ep}\Delta G_{Mp}^*+ \right .\nonumber \\ 
&&\left .\sqrt{\tau (\tau -1)}\cos\theta G_{Mp}F_{3p}^* \right ). 
\label{eq:eq18}
\end{eqnarray}

One can see that:

- The coefficients $D_{xx}, $ $D_{yy}, $ $D_{zz},$ $D_{xz},$ and $D_{zx}$ 
are T--even observables, whereas the coefficients $D_{xy}$, $D_{yx}$, $D_{yz}$,  and $D_{zy}$ are  T--odd observables.

- In the Born approximation the expressions for the T--even correlation 
coefficients coincide with the results of Ref. \cite{Bi93}. The expressions 
for the T--odd ones coincide with the corresponding components of the
polarization correlation tensor of the baryon $B$ and the antibaryon $\bar B$ 
created through the one--photon--exchange mechanism in the $e^+e^-\to B\bar B$ 
process \cite{Du96}.

- The relative contribution of the interference terms (between one- and 
two--photon--exchange mechanisms) increases as $q^2$ becomes larger 
(see the discussion above).

At the reaction threshold the correlation coefficients have some specific
properties:

   - All correlation coefficients do not depend on the function $F_3$.
   
   - In the Born approximation the $(D_{xx}+D_{yy}+D_{zz})$ observable does not depend on the $\theta $ variable, but the TPE contribution induces such dependence.
     
   - In the Born approximation the $D_{yy}$ observable is zero, but the inclusion of the TPE term leads to a nonzero value, determined by the quantity $ReG_M(\Delta G_E-\Delta G_M)^*.$
     
   - The relation $D_{yy}+D_{zz}=0$ holds for $\theta = 90^0$.
   
   - All T--odd double-spin observables vanish.

Taking into account the P--invariance of the hadron electromagnetic interaction,
we can write the following general formula for the differential cross 
section as a function of the polarizations of the proton, (${\vec \xi _1}$), of the 
antiproton (${\vec \xi _2}$) and of the longitudinal polarization of the produced 
lepton, ($\lambda _e$):
\begin{eqnarray}
\displaystyle\frac{d\sigma }{d\Omega}({\vec \xi _1}, {\vec \xi _2}, {\vec \xi })&=&
\left (\displaystyle\frac{d\sigma}{d\Omega}\right)_0
\biggl\{1+A_n{\vec n}\cdot {\vec \xi _1}+\bar A_n{\vec n}\cdot {\vec \xi _2}+
P_n^{(e)}{\vec n}\cdot {\vec \xi }+D_{mm}{\vec m}\cdot {\vec \xi _1}
{\vec m}\cdot {\vec \xi _2}+\nonumber \\ 
&&
D_{nn}{\vec n}\cdot {\vec \xi _1}{\vec n}\cdot {\vec \xi _2}
+D_{{\ell}{\ell}}{\vec {\ell}}\cdot {\vec \xi _1}{\vec {\ell}}\cdot {\vec \xi _2}
+D_{m{\ell}}{\vec m}\cdot {\vec \xi _1}{\vec {\ell}}\cdot {\vec \xi _2}
+D_{{\ell}m}{\vec {\ell}}\cdot {\vec \xi _1}{\vec m}\cdot {\vec \xi _2}+
\nonumber \\ 
&&\lambda _e\biggl [A_m{\vec m}\cdot {\vec \xi _1}
+\bar A_m{\vec m}\cdot {\vec \xi _2}
+A_{\ell}{\vec {\ell}}\cdot {\vec \xi _1}
+\bar A_{\ell}{\vec {\ell}}\cdot {\vec \xi _2}+D_{mn}{\vec m}\cdot {\vec \xi _1}{\vec n}\cdot {\vec \xi _2}+\nonumber \\ 
&&
D_{nm}{\vec n}\cdot {\vec \xi _1}{\vec m}\cdot {\vec \xi _2}
+D_{{\ell}n}{\vec {\ell}}\cdot {\vec \xi _1}{\vec n}\cdot {\vec \xi _2}
+D_{n{\ell}}{\vec n}\cdot {\vec \xi _1}{\vec {\ell}}\cdot {\vec \xi _2}
\biggr ]\biggr\}, 
\label{eq:eq19}
\end{eqnarray}
where 
$${\vec {\ell}}=\frac{{\vec p}}{|{\vec p}|}, \ \ 
{\vec n}=\frac{{\vec p}\times {\vec k}}{|{\vec p}\times {\vec k}|}, \ \ 
{\vec m}={\vec n}\times {\vec {\ell}}, $$
${\vec \xi}$ is the electron polarization four--vector 
and all polarization observables are  functions of two independent variables 
$q^2$ and $\cos\theta .$ The function $A_n$ $(\bar A_n)$ is the asymmetry in the $\bar p+{\vec p}$ $({\vec {\bar p}}+p)$ collision induced by the component of the polarization ${\vec \xi _1}({\vec \xi _2})$ in the direction ${\vec n}$; 
$A_m$ and $A_{\ell}$ ($\bar A_m$ and $\bar A_{\ell}$) are the polarization transfer 
coefficients when the target (beam) and the electron are polarized due to the 
component of polarization vector ${\vec \xi _1} ({\vec \xi _2})$ in the
directions ${\vec m}$ and ${\vec {\ell}}$, correspondingly; $D_{ij}$ ($ij=mm, {\ell}{\ell}, 
nn, m{\ell}, {\ell}m$) and $D_{ij}$ ($ij=mn, nm, n{\ell}, {\ell}n$) are the spin correlation 
coefficients induced by the collision of both polarized initial particles 
for the case of unpolarized and longitudinally polarized final electron, 
respectively; $P_n^{(e)}$ is the electron polarization in the case of 
unpolarized target and beam.  

The following polarization observables 
$$A_n, \ \bar A_n, \ P_n^{(e)}, \ D_{mn}, \ D_{nm}, \ 
D_{{\ell}n}, \ D_{n{\ell}} $$
are T--odd observables, whereas the other ones are T--even observables. 

In the general case, all these polarization observables are nonzero, 
and their $q^2-$ and 
$\cos\theta -$dependence depends on the dynamics of the process. On the basis of C--invariance it is not possible to predict any definite behavior of these observables.

\section{Conclusions}

We have studied the properties of the annihilation process $\bar p+p\to e^++e^-$ in presence of two photon exchange. We have derived the expressions of the cross section and of all polarization observables in terms of the nucleon electromagnetic FFs and of the amplitudes describing the
TPE mechanism. We have analyzed the properties of these observables in different kinematical conditions.

The reasons of the possible contribution of the two photon contribution at large $q^2$ have been discussed long ago for $e+p\to e+p$ elastic scattering and apply equally well to the crossing channels. The importance of the experimental evidence and of the quantitative determination of TPE is related to the extraction of the electromagnetic FFs from the differential cross section. The simple formalism based on the one-photon mechanism,  becomes much more complicated in presence of TPE.

Note that if the charge of the electron and positron is not detected (the detection is symmetric under interchange of the positron and 
electron), then the interference term between the one- and two--photon--exchange 
channels will not contribute to the differential cross section \cite{Zi62, Pu61, 
Dr58}. This symmetry between the positron and the electron can then be used either 
to eliminate or to make evident the influence of the TPE mechanism on the nucleon 
electromagnetic structure.

This analysis is especially useful in view of the future experiments planned
at the FAIR facility, at GSI \cite{GSI}, where  the first
measurement of the relative phase of the proton magnetic and electric FFs
in the time--like region is planned \cite{RL}. This information can discriminate
strongly between the existing models for the nucleon FFs. This phase can be
most simply measured via single--spin asymmetry in the annihilation
reaction (1) with a transversely polarized target or beam.

\section{Acknowledgments}

We acknowledge Prof. M. P. Rekalo for fundamental contributions and ideas on the two-photon exchange problem and for enlightning discussions.

\end{document}